\documentclass[journal,onecolumn,11pt]{IEEEtran}
\usepackage{subfigure,cite,graphicx,amsmath,amssymb,eufrak,mathrsfs,epsf,epsfig}
\usepackage[usenames]{color}
\usepackage{psfrag}

\usepackage[active]{srcltx}

\usepackage{algorithm}
\usepackage{algpseudocode}

\ifCLASSINFOpdf

\else
\fi

\begin{document}
\newtheorem{ach}{Achievability}
\newtheorem{con}{Converse}
\newtheorem{definition}{Definition}
\newtheorem{theorem}{Theorem}
\newtheorem{lemma}{Lemma}
\newtheorem{example}{Example}
\newtheorem{cor}{Corollary}
\newtheorem{prop}{Proposition}
\newtheorem{conjecture}{Conjecture}
\newtheorem{remark}{Remark}
\title{Online Estimation and Adaptation for Random Access  with Successive  Interference Cancellation}
\author{\IEEEauthorblockN{Sang-Woon Jeon,~\IEEEmembership{Member,~IEEE} and Hu Jin,~\IEEEmembership{Senior Member,~IEEE}
}
\thanks{S.-W. Jeon is with the Department of Military Information Engineering, Andong Hanyang University, Ansan 15588, South Korea (e-mail: sangwoonjeon@hanyang.ac.kr).}%
\thanks{Hu Jin is with the Division of Electrical Engineering, Hanyang University,
Ansan 15588, South Korea (e-mail: hjin@hanyang.ac.kr).}}%
 \maketitle

\begin{abstract}
This paper proposes an adaptive transmission algorithm for slotted random access systems supporting the successive interference cancellation (SIC) at the access point (AP).  When multiple users transmit packets simultaneously in a slot, owing to the SIC technique, the AP is able to decode them through SIC resolve procedures (SRPs), which may occupy multiple consequent slots. While such an SRP could potentially improve the system throughput, how to fully exploit this capability in practical systems is still questionable. In particular, the number of active users contending for  the channel varies over time which complicates the algorithm design. By fully exploiting the potential of SIC, the proposed algorithm is designed to maximize the system throughput and minimize the access delay. For this purpose, an online estimation is introduced to estimate the number of active users in real-time and controls their transmissions accordingly. It is shown that the throughput of the proposed algorithm can reach up to 0.693 packets/slot under practical assumptions, which is the first result achieving the throughput limit proved by Yu--Giannakis. It is further shown that the system throughput of 0.559 packets/slot (80.6$\%$ of the throughput limit) is still achievable when the SIC capability is restricted by two.
\end{abstract}
 \IEEEpeerreviewmaketitle



\section{Introduction}  \label{intro}
For various Internet-of-Things (IoT) applications, a massive number of devices are expected to access wireless networks via random access or  machine-to-machine (M2M) communication protocols \cite{mmimo,nbiot}.
In order to support such rapidly growing demands, an efficient random access protocol providing enhanced throughput should be essentially developed, which is operated in a distributed and decentralized manner to cope with excessive uplink accesses for future wireless networks \cite{urllc, vanet}.

Traditionally, it has been well known that the tree (or splitting) algorithm \cite{Cap} and its variants \cite{bert, Mos, Fay} provide improved throughputs. In particular, the first-come first-serve (FCFS) splitting algorithm \cite{bert} has been known as the best one so far, which yields the maximum throughput of  $0.487$ (packets/slot).  
Splitting algorithms in \cite{Cap, bert,Mos,Fay} have been adopted for various wireless systems \cite{Jinn,Yu,Wang, infocom, Kim2017,xu_tree}, as the physical layer technologies have evolved. They have been particularly considered for a cellular system with smart antennas \cite{Jinn} and with successive interference cancellation (SIC) \cite{Yu, Wang}. In particular, for the systems with SIC, the splitting algorithms have been further investigated with several power control schemes incorporated in \cite{Gore}. Note that in general, when more than one packets can be decoded in a slot with advanced signal processing techniques, it is called multipacket reception (MPR) capable channel and thus it encompasses the systems discussed in \cite{Jinn,Yu,Wang,Gore}.

In principle, SIC-enabled techniques can provide substantial throughput improvement of random access systems by utilizing collided packets in the decoding procedure after subtracting interference from decoded packets in a sequential manner.
Because of such potential improvement, SIC techniques have been widely adopted and studied in the literature \cite{Yu,Hu09,Mollanoori,Paolini, Lee}.
A tree-based splitting algorithm has been developed in \cite{Yu} in order to resolve collision efficiently by the help of a SIC-enabled receiver and its modified algorithms have been proposed robust to error-prone networks \cite{Wang,Wang2}, achieving the maximum throughput of $0.693$ (packets/slot).
By allowing more refined feedback for collision resolve beyond tree-based splitting algorithms or allowing a partially centralized control or cooperation between users \cite{Hu09,Mollanoori,Paolini, Lee}, it was shown that the maximum throughput can reach up to one, depending on the types of feedback information and network environment \cite{Paolini,Lee}.    
Decentralized power control and allocation methods have been proposed in \cite{Xu,Lin,Li2} for random access with capabilities of MRP and SIC in order to maximize the sum rate of random access systems by optimally allocating the transmit power of packets and their transmission rates.
    
Recently, more flexible and generalized random access or multiple access frameworks beyond the traditional slot-based user control paradigm have been considered to support massive M2M or IoT traffic demands in an efficient manner \cite{Shirvanimoghaddam}.
Coded random access incorporating with packet erasure codes or rateless-type codes with SIC decoding has been designed in \cite{Paolini2, Stefanovic,Lu}, which provide flexibility and robustness for network dynamics such as fading, bursty noise, and user interference, yielding improved throughput compared to the slotted random access systems.
More recently, non-orthogonal random access (NOMA) has been studied as a promising solution to improve the spectral efficiency for future cellular systems, e.g., fifth generation (5G) systems \cite{Zhang,Ling,Hicuchi}.  For various different types of NOMA systems, SIC is the most well-known detection techniques and actively studied to boost the spectral efficiency in NOMA systems \cite{Zhang,Ling,Hicuchi}.

In spite of recently developed random access or multiple access frameworks and advanced physical layer coding schemes, in this paper, we focus on slotted random access systems with SIC decoding and tree-based splitting algorithm to revolve collision via simple feedback information. It is worthwhile to emphasize that the maximum throughput of $0.693$ (packets/slot) in \cite{Yu} is still quite challenging to achieve in practical environment. There might be three major difficulties to be addressed: 1) SIC capabilities can be limited in practice due to imperfect physical layer processing or excessive complexity; 2) the backlog size is hard to obtain in real-time, which is needed for optimal transmission control; 3) the event of SIC failure can occur because of channel estimation error.
Therefore, the primary aim of this paper to address those constraints and provide throughput close to 0.693 (packets/slot) in more practical random access environment.
Our main contributions  are summarized as follows:
\begin{itemize}
\item A universal online learning method for estimating the current backlog size inspired by the concept of Bayes' rule and an adaptive transmission control algorithm operated with the estimated backlog size are proposed.
\item A novel analytical methodology to derive the maximum throughput limit and the related optimal transmission probabilities are presented, considering practical constraints of bounded SIC capabilities and SIC failure events.  
\item A practical random access system with SIC incorporated with a tree-based splitting algorithm is proposed, achieving the maximum throughput limit by optimally controlling and adapting its transmission under bounded SIC capabilities and SIC failure events.
\end{itemize}

Even though we present the proposed online learning and adaptation method in a standard slotted random access system, the developed online learning framework proposed in this paper is universally applicable for other types of random access systems. 

\section{System Model} \label{sec:system}
We consider a wireless communication system in which an access point (AP) is located at the center of  a cell and time is slotted of equal length for one packet transmission. Users arrive to the system randomly and each user is assumed to have one packet to transmit to the AP. Therefore, a packet and a user are indistinguishable and they are used interchangeably in this paper. Such a system scenario was frequently discussed in the literature and  is also suitable for the machine type communications to support internet of things (IoT) where most of the  machine devices transmit one small packet to the AP. It is worthy to note that we \emph{do not restrict the traffic arrival to be Poisson} which is frequently assumed in the literature on random access systems. Due to the random arrival of the users, the number of active users in the system, i.e., the backlog size, also varies over time. Such a time-varying backlog size is the main difficulty of designing efficient protocols for random access systems.

The backlogged users \emph{randomly} transmit their packets with a certain probability broadcasted by the AP and consequently the number of packets arrived at the AP also varies over time.  When a single packet is received at the AP, it can decode the packet successfully. On the other hand, if the AP receives multiple packets simultaneously at a slot, it is possible to perform SIC for decoding. Considering the difficulty of implementing SIC practically at the physical layer, in this paper we abstract the  \emph{SIC capability} by $M$. That is, if there are more than $M$ users' simultaneous transmissions, the AP is not able to perform SIC due to some technical limitations of physical layer decoding and this slot is denoted as a collision slot. Alternatively, if the number of transmitting users $k$ is between $[2: M]$, the AP initiates an \emph{SIC resolve procedure (SRP)}, denoted by \emph{SRP}$(k)$, from the next slot. Within \emph{SRP}$(k)$, only the $k$ transmitted users are allowed to retransmit their packets under the  \emph{SIC-enabled collision resolve algorithm} which will be introduced in Section \ref{sec:CRA}. Note that when initiating \emph{SRP}$(k)$, the AP is aware of the  $k$ value which is the minimum information of performing SIC. With this information,  \emph{SRP}$(k)$ can be terminated when all the $k$ packets are successfully received. 
We denote $X_k$ $(k \in [2:M])$ by the duration for \emph{SRP}$(k)$ and $E[X_k]$ by its mean. For differentiation, we define the slots that are not in any SPR by the \emph{normal} slots. 
We assume that the AP provides the feedback information of success, collision and, initiating and terminating  the SRP at the end of each slot instantaneously. 

%
%

\begin{figure*}
\centering
\includegraphics[width = 7in]{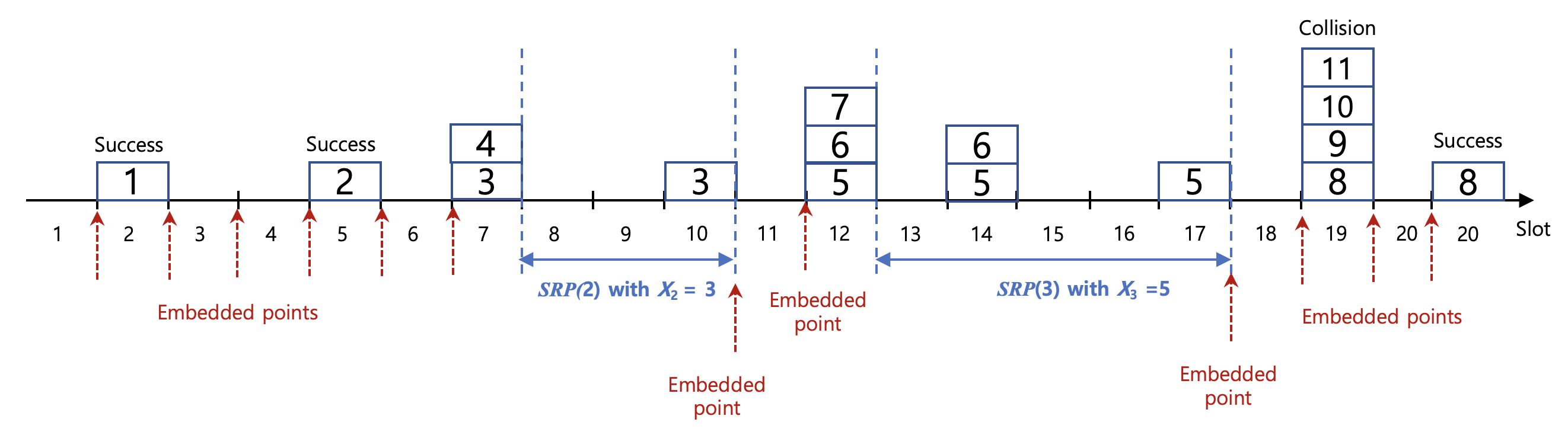}
\caption{Example of transmission behavior over time when $M=3$. } \label{fig:system_model}
\end{figure*}

Fig. \ref{fig:system_model} shows an example of packet transmissions over time when the SIC capability is $M=3$. At Slots 2, 5, and 20, only one packet is transmitted which would be successfully received by the AP. At Slot 7, two packets are transmitted and the AP initiates an \emph{SRP}$(2)$ which lasts for 3 slots, i.e., $X_2 = 3$. In terms of decoding at the AP, as it receives only \emph{Packet 3} at Slot 10 and it is able to decode \emph{Packet 4} by performing SIC for the signal received previously at Slot 7. How those packets are retransmitted in the \emph{SRP}$(2)$ will be detailed in Section \ref{sec:CRA}. Similarly, at Slot 13, an \emph{SRP}$(3)$ is initiated and, in this case, the duration $T_3$ is 5. Finally, at Slot 19, as there are more than $M (=3)$ users' transmissions, a collision occurs and the AP cannot initiate an SRP. We define the \emph{embedded points} as the end of each idle slot, success slot, collision slot and the last slot of each SRP, while we do not count the idle and success slots occurred within any SRP for the embedded points. As a representative example, in Fig. 1, all the embedded points are tagged. The embedded points are always located in front of the \emph{normal} slots. 

For such a system, our primary purpose is to optimally control the transmissions of the backlogged users so that the system could provide maximum throughput as well as minimum access delay for each  user. Note that as the communication of the system bases on the random access, the AP has no information about users, such as  who have packets to transmit  and  what the current backlog size is. To achieve such a goal, we design a transmission control algorithm by clearing the following missions one by one. 
\begin{enumerate}
\item How to design the SIC-enabled collision resolve algorithm so that the duration of \emph{SRP}$(k)$ is minimized for each $k (\in [2:M])$. Note that the users involved in \emph{SRP}$(k)$ should transmit their packets in a distributed manner, i.e., the AP cannot schedule specific user's transmission as it has no information about the users who have packets. 

\item How to control the transmissions of the backlogged users in \emph{normal} slots (other than the slots in an SRP) so that the system operates optimally while exploiting the potentials of the SIC-enabled collision resolve algorithm? 
\end{enumerate}

We should comment here that the random access systems with SIC was also analyzed in \cite{Yu} and it showed that the maximum throughput could reach 0.693. However, \cite{Yu} considered an extreme scenario of \emph{unbounded} SIC capability of $M=\infty$, i.e., no matter how many packets that are  transmitted simultaneously, the AP is always able to perform SIC. This assumption also simplifies the optimal transmission control of the backlogged users: whenever an SRP is completed, all the backlogged users if exist in the system should transmit with probability one (which is called \emph{Gated Access} in \cite{Yu}). Unfortunately, such an operation no longer the optimal for the random access systems with \emph{bounded} SIC capability $M$. For example, if the backlog size is larger than $M$ at a slot, the transmission probability of one would always result in collisions in the consequent slots. Therefore, alternative transmission control algorithm should be designed when it comes to the bounded SIC capability.

We shall design the SIC-enabled collision resolve algorithm in Section \ref{sec:CRA} to solve the above mentioned first mission. In Section \ref{sec:ideal}, we start our discussion by assuming that the backlog size is known at each normal slot and investigate the optimal system performance. This discussion further motivates us to design a transmission control  algorithm that can estimate the the backlog size online while also accommodating the advantages of the SIC-enabled collision resolve algorithm in Sections \ref{sec:proposed_1}.

\section{System Optimization Assuming Known Backlog Size} \label{sec:ideal}
We first investigate an \emph{ideal} scenario where the backlog size $n$ is assumed to be known at each normal slot. In practice, such information cannot obtained in practice as the traffic arrivals are random. Investigation on such an ideal scenario enables us to observe a performance upper bound and also gives us useful insight for designing a transmission control algorithm. 

Suppose that there are $n$ backlogged users in a normal slot and they transmit their packet with probability $p$. Then, according to Renewal Theory, the expected service rate can be derived as 
\begin{align} \label{eq:S_n_M}
R_{n, M}(p) &= \frac{B_n^p(1) + \sum_{k=2}^M k B_n^p(k) }{1 - \sum_{k=2}^M   B_{n}^p(k)+ \sum_{k=2}^M   B_{n}^p(k) ( 1 + E[X_k]) } \notag \\
&= \frac{\sum_{k=1}^M k B_n^p(k) }{1 + \sum_{k=2}^M   B_{n}^p(k)  E[X_k] },
\end{align}
where ${B}_{n}^p(k) = {n \choose k} p^k(1-p)^{n-k}$ denotes the binomial distribution with parameters $p$ and $n$.  The numerator indicates the average number of packets successfully transmitted between two consecutive embedded points and the denominator indicates the corresponding duration.

%
%
%
%

Once the value of $n$ is known, based on \eqref{eq:S_n_M}, we can find the optimal transmission probability $p$ that maximizes the service rate. If such an optimal probability could be controlled slot by slot, we can maximize the system throughput and minimize each user's access delay. While we can optimize the system operation if the backlog size is known prior, it is hard to obtain such information in practice. Motivated by this, in the next Section \ref{sec:proposed_1},  we propose a transmission control algorithm that has the capability of estimating the backlog size online. It is worthy to highlight that online estimation allows the algorithm to adapt to network dynamics.

\section{Proposed Transmission Control Algorithm} \label{sec:proposed_1}


In order to estimate the backlog size in each slot, we adopt the concept of Bayes' rule. As well known, based on \emph{a priori} distribution, Bayes' rule introduces \emph{a posteriori} distribution given that an event is observed. Therefore, when estimating the backlog size, instead of giving an constant number for the estimated backlog size, Bayes' rule estimates it through introducing a distribution. If we define the observed event as the number of users transmitted at each slot, we can update the estimated distribution of the backlog size slot by slot recursively. That is, the \emph{a posteriori} distribution of the current slot becomes the a priori distribution of the next slot. Unfortunately, to implement such a process, we need to calculate the distribution at each slot which accompanies high computational complexity. In order to reduce the computational burden that occurs at each slot, we assume the backlog size follows the Poisson distribution, which is uniquely characterized by its mean. Then, by simply updating the mean we can update its distribution. Such an assumption was also made in \cite{Rivest1987} and was shown to be effective when applying to random access systems.

Let the distribution of the backlog size $n$ to be Poisson with mean $\nu$, i.e,
\begin{equation}
\Phi_\nu(n) = \frac{\nu^n}{n!} e^{-\nu}.
\end{equation}
Then, for the estimation of the backlog size, we need to update $\nu$ recursively slot by slot.

\subsection{Optimal Transmission Probability} \label{subsec:transmission_prob}
For a given backlog size $n$, the service rate is given in \eqref{eq:S_n_M}. If $n$ is a random variable with distribution $\Phi_\nu(n)$, the average service rate is expressed as 
\begin{align}
\hat{S}_{\nu, M} (p) &=  \frac{\sum_{n=0}^\infty\sum_{k=1}^M k B_n^p(k) \Phi_\nu(n) } 
{ \sum_{n=0}^\infty\left[1 + \sum_{k=2}^M   B_{n}^p(k)  E[X_k] \right]  \Phi_\nu(n)} \notag \\
&=  \frac{\sum_{k=1}^M k \Phi_{p\nu}(k) } 
{ 1 +  \sum_{k=2}^M   E[X_k]  \Phi_{p\nu}(k)} \notag \\
&=  \frac{\sum_{k=1}^M k \Phi_{x}(k) } 
{ 1 +  \sum_{k=2}^M   E[X_k]  \Phi_{x}(k)} ,
\end{align}
where we have applied the fact that $\sum_{n=0}^\infty B_n^p(k)\Phi_{\nu}(n) = \Phi_{p\nu}(k)$ and $x =p \nu$ denotes the average number of transmissions per slot. Once $M$ and $E[X_k]$ are given, we can numerically search for the optimal $x^*_M$ that maximizes the above service rate $\hat{S}_{\nu, M} (p) $. Table \ref{TB_S} summarizes the optimal $x^*_M$ and the corresponding maximum service rate $\hat{S}^*_M$, where the values of $E[X_k]$ are obtained from Section \ref{sec:CRA} . Consequently, the optimal transmission probability for a given mean backlog size $\nu$ is derived as 
\begin{equation}\label{eq:p_opt}
p^* = \frac{x_M^*}{\nu}. 
\end{equation}

\begin{table}
\centering
\caption{Optimal parameter values.}  \label{TB_S}
\begin{tabular}{c|c|c|c}
\hline
\hline
$M$ & $x_M^*$ & $\hat{S}^*_M$ & $C_M$\\
\hline\hline
1& 1 & 0.3678 & 1.3922\\
2 & 1.378 & 0.5586 & 2.0458\\
3 & 1.739 & 0.6352 & 2.7020\\
4 & 2.060 & 0.6665 & 3.3833 \\
5 & 2.3762 & 0.6802 & 4.0681\\
10 & 3.8734 & 0.6926 & 7.5626\\
\hline
\end{tabular} 
\end{table}

From Table \ref{TB_S}, we observe that if the SIC capability of 2 or 3 can be obtained, the service rate of 0.5586  or 0.6352 is achievable, which corresponds to 80.6\% and 91.6\% of the maximum throughput  of 0.693 obtained with unbounded SIC capability \cite{Yu}. Therefore, from the engineering viewpoint, one can expect that implementing the SIC capability of 3 at the physical layer is reasonably good. It is notable that higher SIC capability requires more complicated  advanced signal processing techniques, which in general accompany higher implementation complexity and cost.

\subsection{Estimation on the Backlog Size}
Suppose that the transmission probability is $p$. Then, the probability that $m$ among $n$ users are transmitting is obtained as 
\begin{equation}
\Pr(m |n) = B_n^p(m), ~~~~\text{for } m \in [0:n].
\end{equation}
Moreover, given the \emph{a priori} distribution of $\Phi_n(\nu)$, the unconditional probability of $m$ transmitting users is be derived as 
\begin{align}
\Pr(m) & = \sum_{n=0}^\infty  B_n^p(m) \Phi_\nu(n)\notag \\
		  &= \sum_{n=0}^\infty  \Phi_{p\nu}(m) \Phi_{(1-p)\nu}(n-m) \notag \\
		  & = \Phi_{p\nu}(m).
\end{align}
If there are $m$ users' simultaneous transmissions in a slot, by applying Bayes' rule, we obtain
\begin{align}
\Pr(n | m) &= \frac{\Pr(m | n) \Phi_\nu(n)}{ \Pr(m)} \notag \\
	& = \frac{B_n^p(m) \Phi_\nu(n) }{\Phi_{p\nu}(m)} = \Phi_{(1-p)\nu}(n-m),
\end{align}
which is the \emph{a posteriori} distribution of the backlog size $n$. Interestingly, it is still Poisson but with an updated mean by 
\begin{equation}
E[n | m] = (1-p) \nu + m. 
\end{equation}
If the optimal transmission probability $p^* = x^*_M / \nu$  shown in \eqref{eq:p_opt} is applied, we have 
\begin{equation}
E[n | m] = \nu - x^*_M + m,
\end{equation}
where $x^*_M$ is given in Table \ref{TB_S}. Now we are ready to discuss different channel observations such as idle, success,  SRP,  and collision. 

\subsubsection{Idle event} An idle event occurs when $m=0$. Then, the mean backlog size of the \emph{a posteriori} distribution is obtained as 
\begin{equation}
E[n | \text{Idle}] = \nu - x^*_M.
\end{equation}

\subsubsection{Success event} A success event occurs when $m=1$. Consequently, the mean backlog size of the \emph{a posteriori} distribution is 
\begin{equation}
E[n | \text{Success}] = \nu - x^*_M + 1.
\end{equation}
Since one packet is successfully received by the at the AP after a success event, the estimation on the mean backlog size should be further subtracted by 1.


\subsubsection{SRP event} When $m \in [2: M]$, the AP initiates an \emph{SRP}$(m)$. For such an event, the mean backlog size of the \emph{a posteriori} distribution is obtained as 
\begin{equation}
E[n | \text{\emph{SRP}$(m)$}] = \nu - x^*_M +m.
\end{equation}
Since those $m$ packet would be eventually received by the AP at the end of the \emph{SRP}$(m)$, the estimation on the mean backlog size should be further subtracted by $m$ at the end of the SRP.

%
%
%
%
%
%
%
%
%
\subsubsection{Collision event}
When $m > M$, a collision occurs and the AP is not able to know the number of users involved in the collision. The probability that a collision occurs when there are $n$ backlogged users in the system is obtained as 
\begin{equation}
\Pr(C|n) = \sum_{m=M+1}^n B_n^p(m).
\end{equation}
Consequently, the unconditional  probability of a collision event is obtained as 
\begin{align}
\Pr(C) & = \sum_{n = M}^\infty  \left( \sum_{m=M+1}^n B_n^p(m) \right) \Phi_\nu(n)\notag \\
 	& =   \sum_{m=M+1}^\infty  \Phi_{p\nu}(m) \notag \\
	& =  1 - \sum_{m=0}^M  \Phi_{p\nu}(m).
\end{align}
Then,  by applying Bayes' rule, we obtain the conditional probability of $n$ given a collision event by
\begin{align}
\Pr(n | C) &= \frac{\Pr(C|n) \Pr(n)}{\Pr(C)} \notag \\ 
			& = \frac{ \Phi_n(\nu) \sum_{m=M+1}^n B_n^p(m)}{  1 - \sum_{m=0}^M  \Phi_{p\nu}(m)} \notag \\
			& = \frac{  \sum_{m=M+1}^n \Phi_{(1-p)\nu}(n-m) \Phi_{p\nu}(m)}{  1 - \sum_{m=0}^M  \Phi_{p\nu}(m)}.
\end{align}
It is not difficult to check that the above distribution is no longer the Poisson while we still approximate the distribution by Poisson with its mean as 
\begin{align}
E[n | C] & = \sum_{n=0}^\infty n \Pr(n | C) \notag \\
	& = (1 - p) \nu + \frac{p\nu - \sum_{m=0}^M m\Phi_{p\nu}(m)}{1 - \sum_{m=0}^M  \Phi_{p\nu}(m)}.	
\end{align}
If the optimal transmission probability of $p^* = x^*_M / \nu$ is applied, we have 
\begin{align}
E[n | C ] &= \nu - x^*_M + \frac{x^*_M - \sum_{m=0}^M m\Phi_{x^*_M}(m)}{1 - \sum_{m=0}^M  \Phi_{x^*_M}(m)} \notag \\
			& = \nu + C_M,
\end{align}
where $C_M = \frac{\sum_{m=0}^M (x^*_M - m)\Phi_{x^*_M}(m) }{ 1 - \sum_{m=0}^M  \Phi_{x^*_M}(m)}$ and its value for each $M$ is listed in Table \ref{TB_S}.

\subsection{Algorithm Description}

Algorithm \ref{alg:limited_SIC_cap} describes the overall procedure of our proposed algorithm. Note that the values of $x^*_M$ and $C_M$ are listed in Table \ref{TB_S}. Lines 3, 6, 11, 14 correspond to the estimation on the arrival rate where the previous arrival rate is given by multiplying the weight of $\theta$. In Section \ref{sec:simul}, we set $\theta=0.99$ in simulation. Lines 4, 7, 12, 15 correspond to the update on the mean backlog size whose mathematical background was introduced in the previous subsection. Line 17 corresponds to the calculation on the optimal transmission probability. 

\begin{algorithm}[pt]\caption{Estimation and Adaptation Algorithm at the AP with the SIC capability $M$} \label{alg:limited_SIC_cap}
	\begin{algorithmic}[1]\label{algb} 
		\State Initialize $\nu = 10$, $ p^* = x^*_M / M$, $\lambda_E = 0.5$ and repeat the following at each embedded point. 
		\If {Idle}
		\State $\lambda_E \leftarrow \theta \lambda_E$		
		\State  $\nu \leftarrow \nu - x^*_M + \lambda_E $  
		\ElsIf {Success}
		\State $\lambda_E \leftarrow \theta \lambda_E + (1-\theta)\cdot 1$ 
		\State $\nu \leftarrow \nu - x^*_M + \lambda_E $  
		\ElsIf	 {\emph{SRP}$(m)$ with $m \in [2: M]$ }
	 	\State 	 Initiates an \emph{SRP}$(m)$ with the involved  $m$ users.
		\State	 Record the SRP duration $X_m$. 		
		\State $\lambda_E \leftarrow \frac{\theta \lambda_E + (1-\theta)        m}{\theta \cdot 1 + (1-\theta) (1 + X_m)}$		
		\State $\nu \leftarrow \nu - x^*_M + \lambda_E T_m$  
     	\ElsIf	 {Collision}
		\State $\lambda_E \leftarrow \theta \lambda_E$		
		\State  $\nu \leftarrow \nu + C_M + \lambda_E $  
     	
 	\EndIf
		\State  $p^* = \min \left(1, \frac{x^*_M}{\nu} \right)$
		\State Broadcast $ p^*$ for the next time slot.
	\end{algorithmic}
\end{algorithm} 

%
%
%
%
%
%
%
%
%
%
%
%
%
%
%

\section{SIC Resolve Algorithm} \label{sec:CRA}
The proposed transmission control algorithm, introduced in the previous Section \ref{sec:proposed_1}, adapts the transmission probabilities for packet transmissions at each \emph{normal} slot. For the design of the proposed algorithm, we applied the results of the average duration of \emph{SRP}$(m)$  which is initiated by $m (\in[2:M])$ users' simultaneous transmissions. Once an \emph{SRP}$(m)$ starts, only the $m$ users involved in the previous transmissions retransmit their packets under the \emph{SIC resolve algorithm} in the subsequent slots. If all the $m$ packets are successfully received by the AP, \emph{SRP}$(m)$ is terminated. Section \ref{subsec:SIC_alg} introduces the SIC resolve algorithm and  Section \ref{subsec:E_X} analyzes the expected duration of each \emph{SRP}$(m)$, i.e., $E[X_m]$.

\subsection{Algorithm Description}\label{subsec:SIC_alg}

Let $\mathcal{V}$ denote the set of the users that make the AP initiate an \emph{SRP}$(|\mathcal{V}|)$ where $|\mathcal{V}|$ denotes the cardinality of the set $\mathcal{V}$. Then, the SRP is proceeded by invoking the function \Call{SIC\_Resolve}{$\mathcal{V}, X_{|\mathcal{V}|}$} recursively, which is described in Algorithm \ref{alg:collision_resolve}. 
$X_{|\mathcal{V}|}$ is an output of the function \Call{SIC\_Resolve}{$\mathcal{V}, X_{|\mathcal{V}|}$} and it indicates the duration spent for the function. 

At the first slot of \Call{SIC\_Resolve}{$\mathcal{V}, X_{|\mathcal{V}|}$}, the users in $\mathcal{V}$ retransmit their packet with probability $p_{|\mathcal{V}|}$ as in Line 6. This retransmission probability could be minimized for each $|\mathcal{V}|$ as the AP knows the value of $|\mathcal{V}|$ when invoking the function. Each user who has transmitted in the first slot considers itself as a member of a new set $\mathcal{W}$ as in Line 7. If $|\mathcal{W}| = 1$, the AP decodes the only transmitted user's packet and additionally performs SIC based on the previously received signals. If the number of non-transmitted users is larger than 2, those users become a member of the set $\mathcal{V}\setminus\mathcal{W}$ as in Line 14. If $|\mathcal{W}|\geq 2$, i.e., the number of transmitted users is larger than 1, the \Call{SIC\_Resolve}{} is invoked for $\mathcal{W}$ as in Line 17. After the completion of \Call{SIC\_Resolve}{$\mathcal{W}, |X_{\mathcal{W}}|$} if exists, the function \Call{SIC\_Resolve}{} runs again for $\mathcal{V}\setminus\mathcal{W}$ if $|\mathcal{V}|-|\mathcal{W}|\geq 2$. The time duration for \Call{SIC\_Resolve}{$\mathcal{V}, X_{|\mathcal{V}|}$} is updated at Lines 8, 18, and 22. 

\begin{algorithm}[pt]\caption{SIC Resolve Algorithm} \label{alg:collision_resolve}
	\begin{algorithmic}[1]\label{alg0}
		\State Initialize $\mathcal{V}$ as the set of users involved.
		\State Run \Call{SIC\_Resolve}{$\mathcal{V}$}.
		
		\Function{SIC\_Resolve($\mathcal{V}$)}{}
		\State Set $\mathcal{W}=\emptyset$ and $X_{|\mathcal{V}|} = 0$.
		\While {$\mathcal{W}=\emptyset$ or $\mathcal{W}=\mathcal{V}$} 
		\State Each user in $\mathcal{V}$ retransmits with probability $p_{|\mathcal{V}|}$.  
 	    \State Each transmitted user becomes a member of  $\mathcal{W}$. 	    
		\State $X_{|\mathcal{V}|} \leftarrow X_{|\mathcal{V}|} + 1$.	
		\EndWhile
		\If {$|\mathcal{W}| = 1$}
		\State The AP performs decoding and SIC as possible.
		\EndIf				
		\If {$|\mathcal{V}|-|\mathcal{W}|\geq 2$}
		\State	Each non-transmitted user enters $\mathcal{V}\setminus\mathcal{W}$.
		\EndIf				
		\If {$|\mathcal{W}|\geq 2$}
		\State Run \Call{SIC\_Resolve}{$\mathcal{W}, |X_{\mathcal{W}}|$}.
		\State $X_{|\mathcal{V}|} \leftarrow X_{|\mathcal{V}|} + X_{|\mathcal{W}|}$.
		\EndIf
		\If {$|\mathcal{V}|-|\mathcal{W}|\geq 2$}
		\State Run \Call{SIC\_Resolve}{$\mathcal{V}\setminus\mathcal{W}, X_{|\mathcal{V}\setminus\mathcal{W}|}$}.
		\State $X_{|\mathcal{V}|} \leftarrow X_{|\mathcal{V}|} + X_{|\mathcal{V}\setminus\mathcal{W}|}$.
		\EndIf
		\EndFunction		
	\end{algorithmic}
\end{algorithm}

\subsection{Expected Duration of SIC Resolving Time} \label{subsec:E_X}
In this subsection, we analyze the expected SIC resolve time achievable by Algorithm \ref{alg:collision_resolve}. 

\subsubsection{Recursive relation for SIC resolving time} \label{subsec:resolve_time}

Let $X_m$ be the number of time slots for SIC resolving when $m$ users are initially involved. 
Obviously, $X_0=X_1=0$.  To derive the recursive relation between $\{X_m\}$, denote $\Delta_m\geq 1$ as the number of time slots required for starting the inherent SIC resolve phase like Lines 17 or 21 in Algorithm \ref{alg:collision_resolve}, 
and $M_m\in[1:m-1]$ as the number of  users for the inherent SIC resolve phase.
Notice that both $\Delta_m$ and $M_m$ are random.

From the proposed SIC resolve algorithm, we have $X_2=\Delta_2$, $X_3=\Delta_3+X_2$, and for $m\geq 4$, 
\begin{align} \label{eq:t_n}
X_m=\begin{cases}
\Delta_m+X_{m-1},~ ~~~~\mbox{if } M_m=1 \mbox{ or }M_m=m-1,\\ \Delta_m+X_{M_m}+X_{m-M_m}, ~~\mbox{if } M_m\in[2:m-2],
\end{cases}
\end{align}

\begin{figure}[t]
\centering
\includegraphics[scale = 0.35]{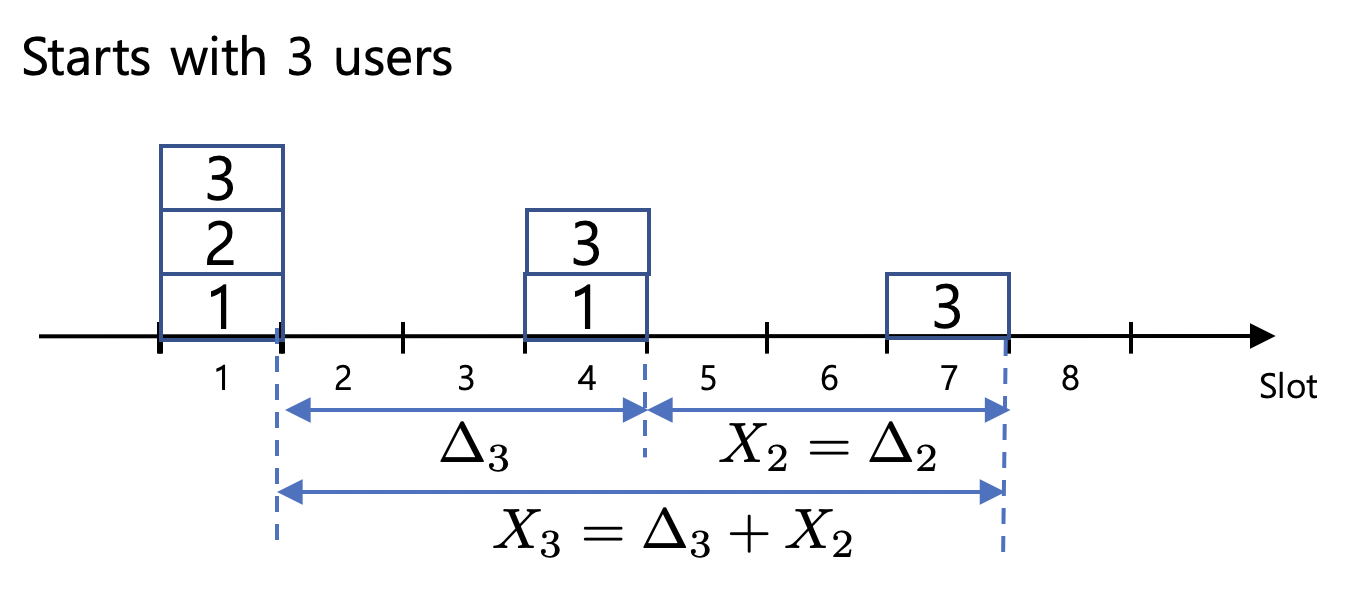}
\caption{Explanatory SIC resolving time when $3$ users are involved.} \label{fig:ex_t_3}
\end{figure}

\begin{figure}[h]
\centering
\includegraphics[scale = 0.35]{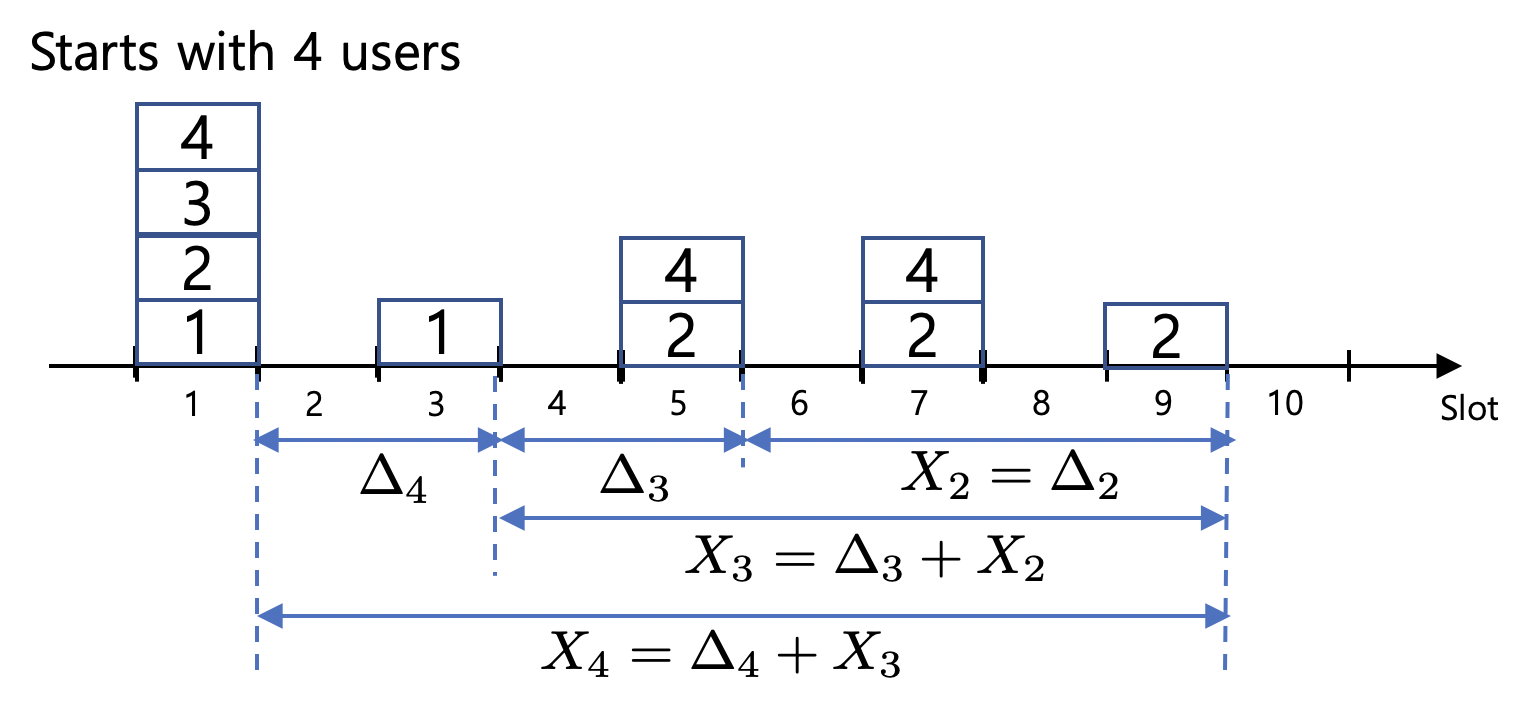}
\caption{Explanatory SIC resolving time when $4$ users are involved.} \label{fig:ex_t_4}
\end{figure}

For better understanding of the relation between Algorithm \ref{alg:collision_resolve} and $X_m$, explanatory cases for resolving SIC started from three and four users are provided in Figs. \ref{fig:ex_t_3} and \ref{fig:ex_t_4}, respectively.
Specifically, in Fig. \ref{fig:ex_t_3}, three users (users 1 to 3) are initially involved and after random transmissions with probability $p_3$, users 1 and 3 retransmit after three time slots, i.e., $X_3=\Delta_3+X_2$ and $\Delta_3=3$ in this case. Note that user 2 does not participate the remaining retransmission procedure because of the condition in Line 20 of Algorithm \ref{alg:collision_resolve} is not satisfied.  
Then users 1 and 3 perform random transmissions with probability $p_2$ and user 3 retransmits alone after three time slots.
Again, from the condition in Line 20 of Algorithm \ref{alg:collision_resolve}, user 1 does not participate the remaining retransmission procedure, which gives $X_2=\Delta_2$ and $\Delta_2=3$ in this case.

In Fig. \ref{fig:ex_t_4}, four users (users 1 to 4) are initially involved and after random transmissions with probability $p_4$, user 1 retransmits after two time slots, i.e., $X_4=\Delta_4+X_3$ and $\Delta_4=2$.
Then users 2, 3, and 4 perform random transmissions with probability $p_3$ and users 2 and 4 retransmit after two time slots.
In this case, $X_3=\Delta_3+X_2$ and $\Delta_3=2$.
Finally, users 2 and 4 perform random transmissions with probability $p_2$ and user 2 alone retransmits after four time slots, i.e., $X_2=\Delta_2$ and $\Delta_2=4$.
Note that users 2 and 4 retransmit after two time slots, but the corresponding received signal is removed at the AP and users 1 and 4 perform the same random transmissions, see the condition in Line 5 of Algorithm \ref{alg:collision_resolve}.

\subsubsection{Expected Duration Analysis} \label{subsec:expected_time}
Now consider how to derive the expected time required for resolving SIC, i.e., $E[X_m]$ for $m\in[2:M]$.

From the definition of $\Delta_m$, it follows the geometric distribution with success probability $1-B^{p_m}_m(0)-B^{p_m}_m(m)$ and, as a result, we have
\begin{align} \label{eq:delta_n}
E[\Delta_m]=\frac{1}{1-B^{p_m}_m(0)-B^{p_m}_m(m)}.
\end{align}
Also, for $l\in[1:m-1]$, we have
\begin{align} \label{eq:m_n}
\Pr(M_m=l)=\frac{B^{p_m}_{m}(l)}{1-B^{p_m}_{m}(0)-B^{p_m}_{m}(m)},
\end{align}
which satisfies $\sum_{l=1}^{m-1} \Pr(M_m=l)=1$.

Therefore, $E[X_2]=E[\Delta_2]$, $E[X_3]=E[\Delta_3]+E[\Delta_2]$, and from \eqref{eq:t_n},
\begin{align} \label{eq:expected_t_n}
&E[X_m]\nonumber\\
&= \sum_{l=1}^{m-1}  \Pr(M_m=l) E[X_m| M_m=l]\nonumber\\
&=(\Pr(M_m=1)+\Pr(M_m=m-1))\left(E[\Delta_m]+E[X_{m-1}]\right)\nonumber\\
&+\sum_{l=2}^{m-2} \Pr(M_m=l)\left(E[\Delta_m]+E[X_{l}]+E[X_{m-l}]\right)\nonumber\\
&= E[\Delta_m]+\sum_{l=2}^{m-1}(\Pr(M_m=l)+\Pr(M_m=m-l))E[X_l]
\end{align}
for $m\geq 4$.
Then, $E[X_m]$ can be recursively calculated  from \eqref{eq:expected_t_n} by using \eqref{eq:delta_n} and \eqref{eq:m_n}.

Recall that $p_m$ is the retransmission probability of each involved user when $m$ users are involved for the SIC resolving, which can be optimized to minimize the expected SIC resolving time.
Let $\{p_i^*\}_{i=1}^m$ be the set of optimal retransmission probabilities minimizing $E[X_m]$.
Denote the minimum expected number of time slots for resolving SIC among $m$ users by  $E^*[X_m]$. Then, it is given by
\begin{align}
E^*[X_m]&=\min_{\{p_i\in(0,1)\}_{i=1}^m}E[X_m]. 
\end{align}

Table \ref{table:expected_t} summarizes $E^*[X_m]$ for $m=2,3,4,5,10$. Through numerical calculations, we also observed that the setting of $p_m=1/2$ for all $m$ yields almost identical expected duration for resolving SIC as Table \ref{table:expected_t}. Therefore, from the engineering viewpoint, setting a constant probability of $1/2$ is a reasonably good choice.

\begin{table} 
\centering
\caption{Minimum expected number of time slots.}  \label{table:expected_t}
\begin{tabular}{c|c}
$m$ & $E^*[X_m]$\\
\hline\hline
2&  2 \\
3 &  3.333 \\
4 &  4.761 \\
5 &  6.210 \\
10 & 13.426 \\
\hline
\end{tabular} 
\end{table}

\section{Impact of SIC Failure} \label{sec:sic_fail}

Although introducing SIC capability in the physical layer decoding can provide substantial throughput improvement in random access systems, SIC is quite challenging in practice, especially as the number of users involving in the SIC procedure increases due to imperfect channel estimation, error propagation during SIC process, etc.
In this section, we will address such practical issues and how to modify the proposed online transmission control algorithm in Section \ref{sec:proposed_1}, minimizing the performance degradation due to SIC failure.

Let $p_e\in[0,1)$ be the probability of SIC failure in each of SIC steps.
For simplicity, we assume that an event of SIC failure independently occurs with probability $p_e$ for each time slot involving more than one user are simultaneously transmitted.\footnote{The same analysis presented in Section \ref{sec:sic_fail} is straightforwardly applicable for the case where the probability of SIC failure depends on the number of simultaneously transmitted users.} 

Obviously, the most efficient way to deal with such an SIC failure event is that the corresponding users involved in the event retransmit again except the users whose packets are already decoded, to construct the same received signal at the AP so that the AP can resume the rest of SIC procedure.

We can apply the same notion of the SIC resolving time to analyze the impact of SIC failure.
Let $X^{(e)}_m$ be the number of time slots for SIC resolving when $m$ users are initially involved counting the retransmission slots due to SIC failure.
Note that the analytical method introduced in Section \ref{subsec:E_X} is able to provide a simple universal tool for dealing with such probabilistic events.
Define $\Delta_e$ as the number of time slots required to have the received signal with no SIC failure at the AP excluding the originally received signal. Then, its mean is given by $E[\Delta_e]=\frac{1}{1-p_e}-1$ from the geometric distribution.
Now define $X'_m=X^{(e)}_m-\Delta_e$, which corresponds to the number of time slots for SIC resolving when $m$ users are initially involved assuming that there is no SIC failure for the received signal involving $m$ users.

Then, similar methods in Section \ref{subsec:E_X} are applicable to derive recursive relations between $\{X^{(e)}_m\}$ and $\{X'_m\}$.
In particular, $X'_2=\Delta_2$ and $X^{(e)}_2=X'_2+\Delta_e$.
Also, 
\begin{align}
X'_3=\begin{cases}
\Delta_3+X'_2,~ ~~~~\mbox{if }M_3=1,\nonumber\\
\Delta_3+X^{(e)}_2,~ ~~~~\mbox{if }M_3=2,
\end{cases}
\end{align}
and $X_3^{(e)}=X'_3+\Delta_e$.
For $m\geq4$,
\begin{align}
X'_m=\begin{cases}
\Delta_m+X'_{m-1}~ ~~~~\mbox{if }M_m=1,\nonumber\\
\Delta_m+X^{(e)}_{M_m}+X'_{m-M_m}~ ~~~~\mbox{if }M_m\in[2:m-2],\nonumber\\
\Delta_m+X^{(e)}_{m-1}~ ~~~~\mbox{if }M_m=m-1,
\end{cases}
\end{align}
and $X^{(e)}_{m}=X'_{m}+\Delta_e$.

Then, from \eqref{eq:delta_n} and \eqref{eq:m_n}, we have $E[X'_2]=E[\Delta_2]$, $E[X_2^{(e)}]=E[X'_2]+E[\Delta_e]$.
Also, 
\begin{align}
E[X'_3]=E[\Delta_3]+E[X_2']+\Pr(M_3=2)E[\Delta_e]
\end{align}
and $E[X^{(e)}]=E[X'_3]+E[\Delta_e]$. For $m\geq 4$, we have
\begin{align}
&E[X'_m]\nonumber\\
&=E[\Delta_m]+\sum_{l=2}^{m-1}(\Pr(M_m=l)+\Pr(M_m=m-l))E[X'_l]\nonumber\\
&+\sum_{l=2}^{m-1}\Pr(M_m=l)E[\Delta_e]
\end{align}
and $E[X_m^{(e)}]=E[X'_m]+E[\Delta_e]$.

Similarly, let $E^*[X^{(e)}_m]$ be the minimum expected number of time slots for resolving SIC among $m$ users. Unlike the case of no SIC failure, $p_m=1/2$ is not optimal if $p_e>0$. For instance, Table \ref{table:expected_t_sic_error} summarizes  $E^*[X^{(e)}_m]$ and the corresponding $\{p_i^*\}_{i=1}^m$ for $m=2,3,4,5,10$ when $p_e=0.5$.

\begin{table} [t!]
\centering
\caption{$E^*[X^{(e)}_m]$ and $p^*_m$ when $p_e=0.5$.}  \label{table:expected_t_sic_error}
\begin{tabular}{c|c|c}
\hline
\hline
$m$ & $E^*[X^{(e)}_m]$ & $p_m^*$ \\
\hline\hline
2&  3 & 0.5  \\
3  & 4.788 & 0.412 \\
4 & 6.633 & 0.343 \\
5 & 8.486 & 0.288\\
10 & 17.802 & 0.163 \\
\hline
\end{tabular} 
\end{table}

Now, from the derived $E^*[X^{(e)}_m]$, we can obtain the maximum service rate and the corresponding parameters required for applying the proposed online transmission control algorithm as the same manner in Section \ref{sec:proposed_1}. 
For instance, Table \ref{table:parameters_sic_fail} summarizes $\hat{S}^*_M$ and the corresponding parameters for online estimation when $p_e=0.5$.


\begin{table}[t!]
\centering
\caption{Parameters for online estimation when $p_e = 0.5$.} \label{table:parameters_sic_fail}
\begin{tabular}{c|c|c|c}
\hline
\hline
$M$ & $\hat{S}^*_M$ & $x_M^*$ & $C_M$\\
\hline\hline
1 & 0.3678 & 1  & 1.3922\\
2 & 0.4821 & 1.2580 & 2.1220\\
3 & 0.5155 & 1.4700 & 2.8892\\
4 & 0.5264 & 1.6380 & 3.6960 \\
5 & 0.5300 & 1.760 & 4.5461\\
10  & 0.5316 & 1.8840 & 9.2964\\
\hline
\end{tabular} \end{table}


\begin{figure}[t] \centering
\includegraphics[scale=0.7]{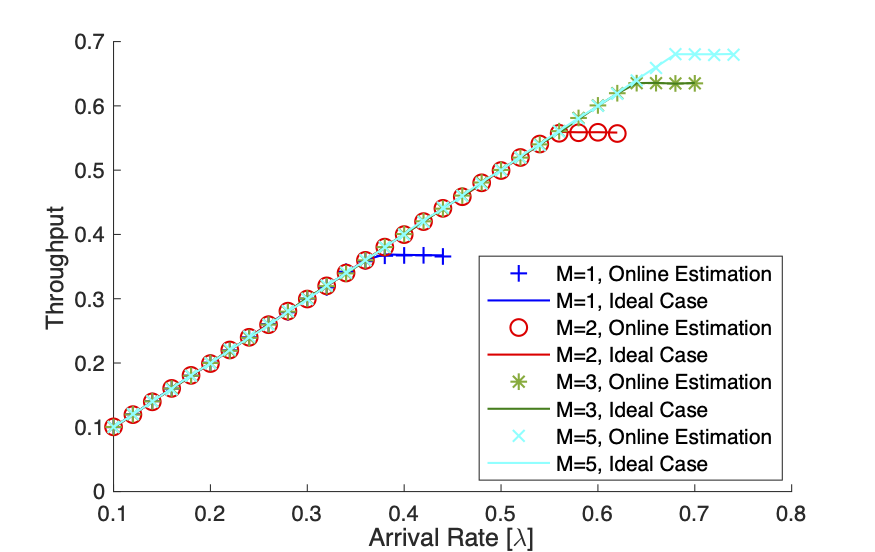}
\caption{Average throughput with respect to $\lambda$.}\label{fig:fig2}
\end{figure}

\begin{figure}[t] \centering
\includegraphics[scale=0.6]{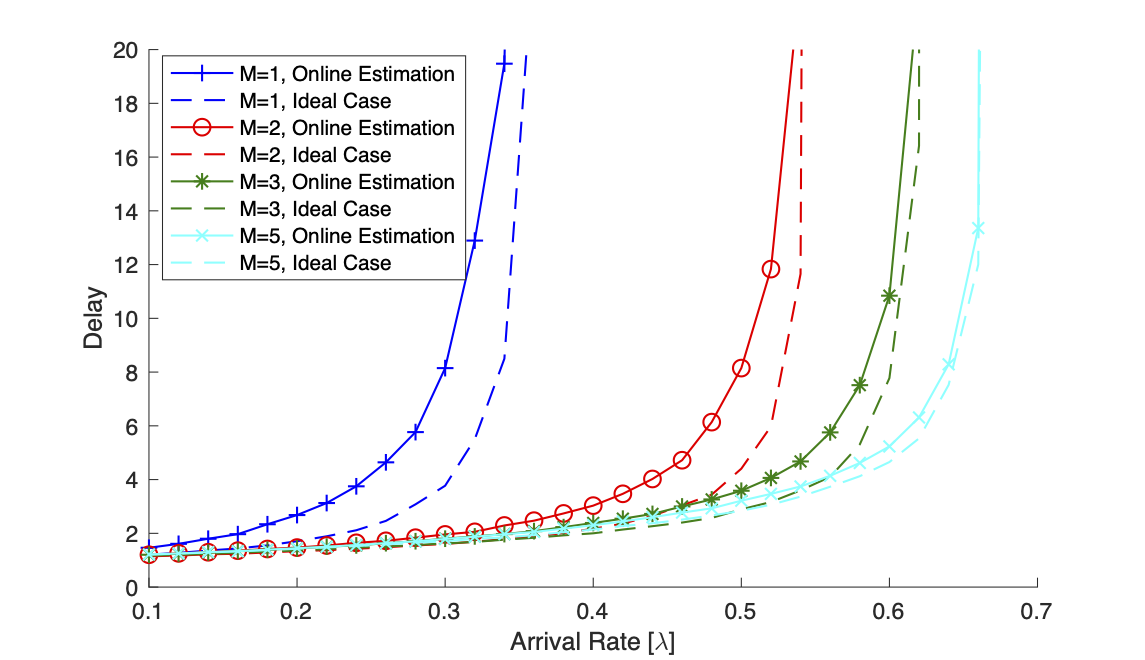}
\caption{Average delay with respect to $\lambda$.}\label{fig:fig3}
\end{figure}

\section{Numerical Evaluation} \label{sec:simul}
In this section, we perform numerical evaluation to demonstrate the performance of the proposed online estimation and transmission control algorithm for random access with SIC.
Recall the throughput improvement in Table \ref{TB_S} requires the backlog size in real-time, which is impossible for most random access systems.
Furthermore, in practice, SIC is quite challenging especially as the number of users involving in the SIC procedure increases due to imperfect channel estimation, error propagation during SIC process, etc.
In the following, we will address these practical issues and demonstrate that the proposed online estimation and adaptation framework can still provide such throughput improvement under more practical environment.
Unless otherwise stated, we assume Poisson packet arrivals and the update weight $\theta$ in Algorithm \ref{alg:limited_SIC_cap} is set as $\theta=0.99$ for the online estimation.

\subsection{Online Estimation and Adaptation}
In order to investigate the performance of the proposed algorithm in Section \ref{sec:proposed_1}, Fig. \ref{fig:fig2} compares the average throughput achievable by the proposed algorithm with that of the ideal case where the backlog size is known.
As seen in the figure, the maximum service rate supportable by the proposed algorithm is the same for that of the ideal case.
In order to investigate the impact of the online estimation and transmission control closely, Fig. \ref{fig:fig3} plots the average access delay with respect to $\lambda$. The average delay rapidly increases  as the arrival rate converges to the maximum service rate so that the system becomes unstable for this regime. Note that the average delay gap between the online estimation and the ideal case is less than four time slots for the stable regime of $\lambda$. The results demonstrate that the proposed online estimation and transmission control performs well by only learning from the historical information.

\begin{figure}[t] \centering
\includegraphics[scale=0.6]{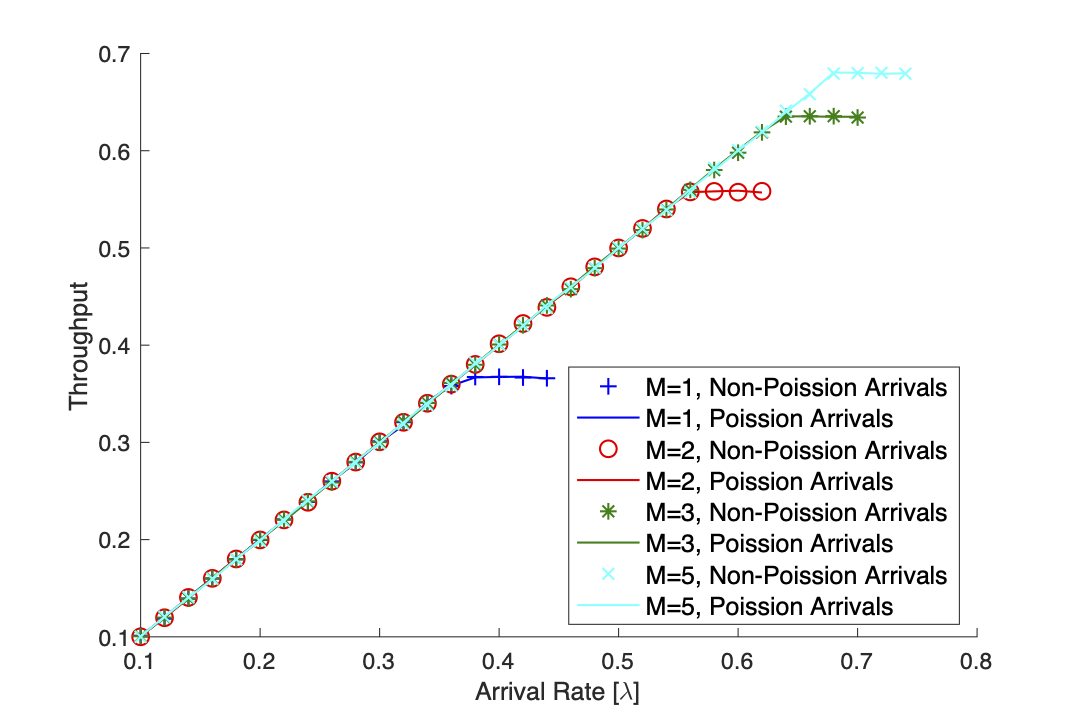}
\caption{Average throughput with respect to $\lambda$ for non-Poisson arrivals.}\label{fig:fig7}
\end{figure}

\begin{figure}[t] \centering
\includegraphics[scale=0.5]{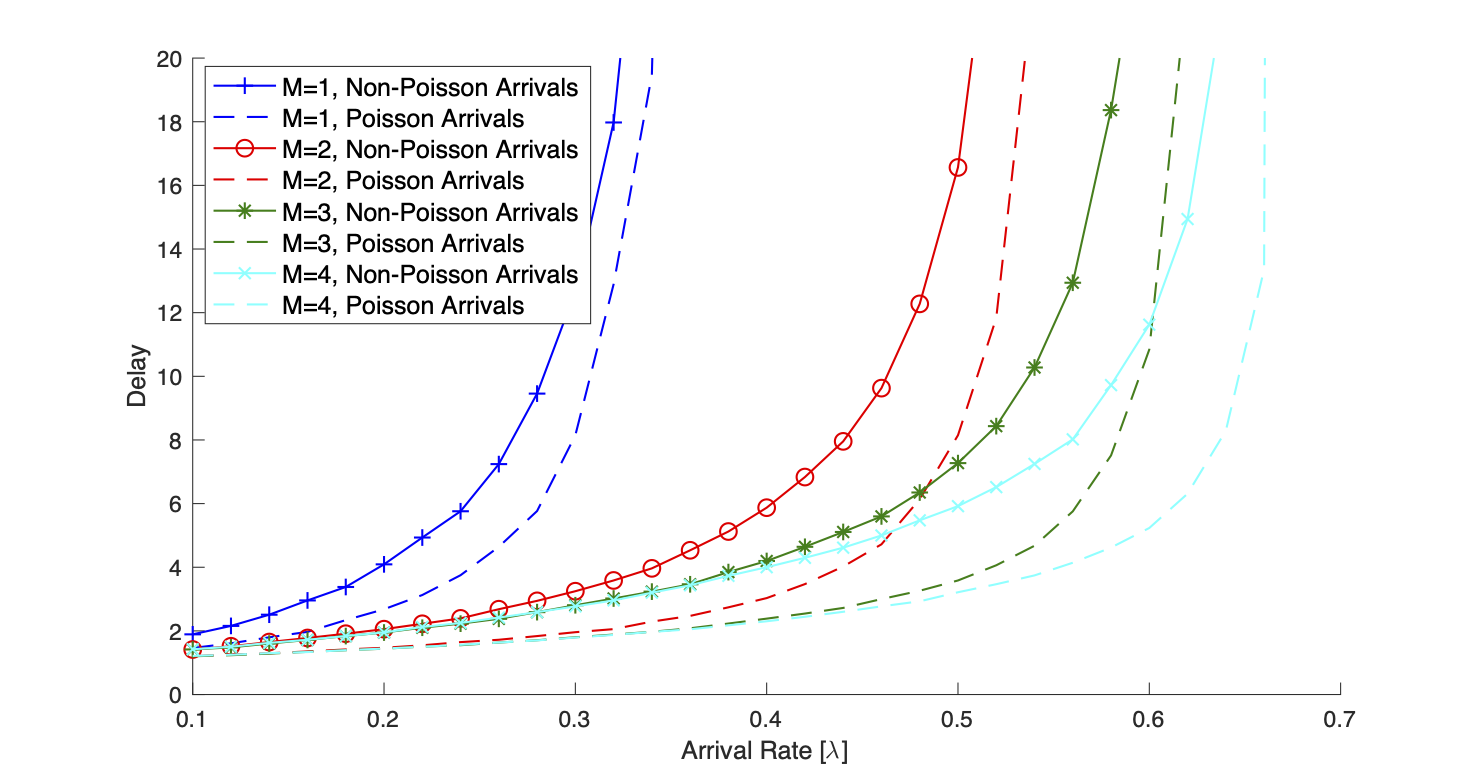}
\caption{Average delay with respect to $\lambda$ for non-Poisson arrivals}\label{fig:fig8}
\end{figure}

Not only for Poisson user arrivals, the proposed algoirthm is universally applicable for a board class of arrival distributions.
As an explanatory case, we assume that for every 100 time slots the arrival rate is changed either the Poisson distribution of mean $2\lambda$ or zero with equal probability. Hence, the mean arrival rate is equal to $\lambda$.
Figs. \ref{fig:fig7} and \ref{fig:fig8}  plot the average throughput and delay of this non-Poisson arrival model, respectively. For comparison, we also plot the performances of the poisson arrival model.
Because the considered non-Poisson model corresponds to on--off user arrivals,  the delay performance is expected to worse that that of the Poisson model.
 As seen in the figures, the proposed algorithm still performs well even for non-Poisson user arrivals.
 
\begin{figure}[t] \centering
\includegraphics[scale=0.55]{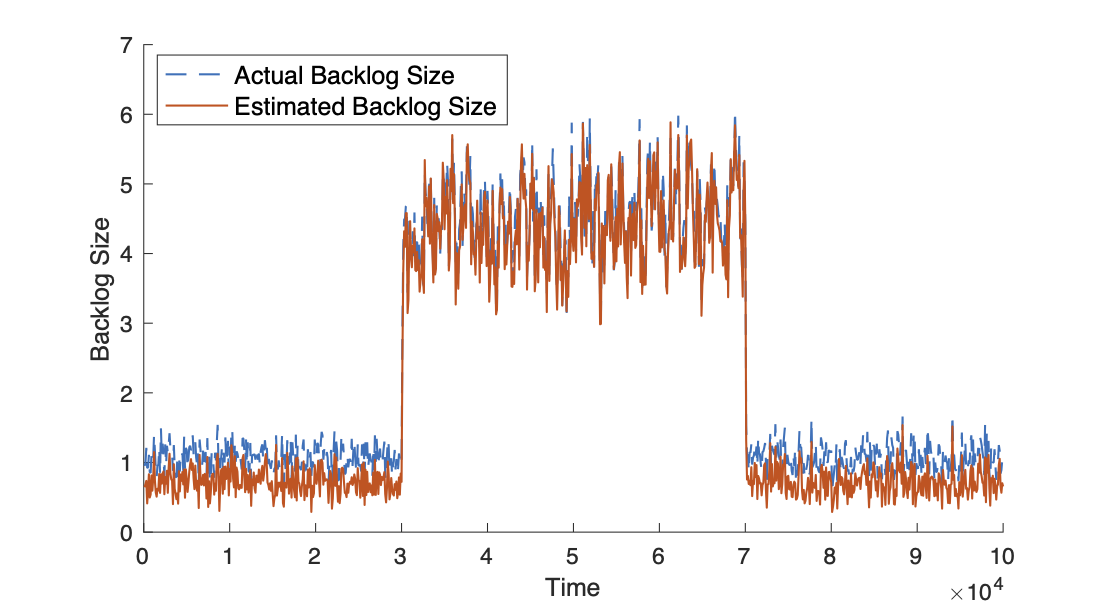}
\caption{Estimated backlog size of Algorithm \ref{alg:limited_SIC_cap} when $M=2$. }\label{fig:fig6}
\end{figure}

To investigate the real-time estimation performance of the proposed algorithm, Fig \ref{fig:fig6} compares the estimated backlog size (the parameter $\nu$ in Algorithm \ref{alg:limited_SIC_cap}) with the actual backlog size by observing $10^5$ time slots when $M=2$, where $100$ time episodes are averaged out to plot the figure. 
For each time episode, we set $\lambda=0.4$ for $t\in[1:3\times 10^4]$ and $t\in[7\times 10^4+1:10^5]$ and $\lambda=0.5$ for $t\in[3\times 10^4+1:7\times10^4]$. The result shows that Algorithm \ref{alg:limited_SIC_cap} can acurately estimate the actual backlog size in real-time, providing an efficient transmission control for random access systems reflecting network dynamics.

\begin{figure}[t!] \centering
\includegraphics[scale=0.7]{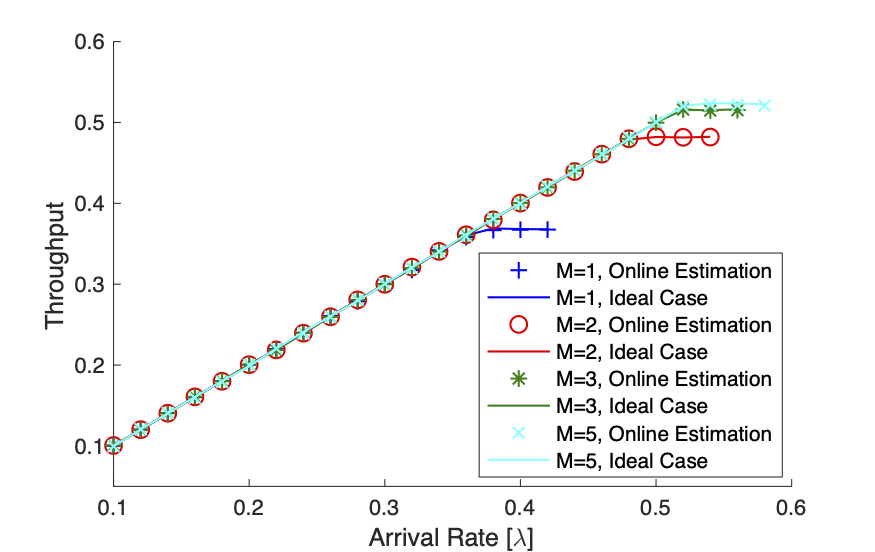}
\caption{Average throughput with respect to $\lambda$ when $p_e=0.5$.}\label{fig:fig4}
\end{figure}

\begin{figure}[t!] \centering
\includegraphics[scale=0.6]{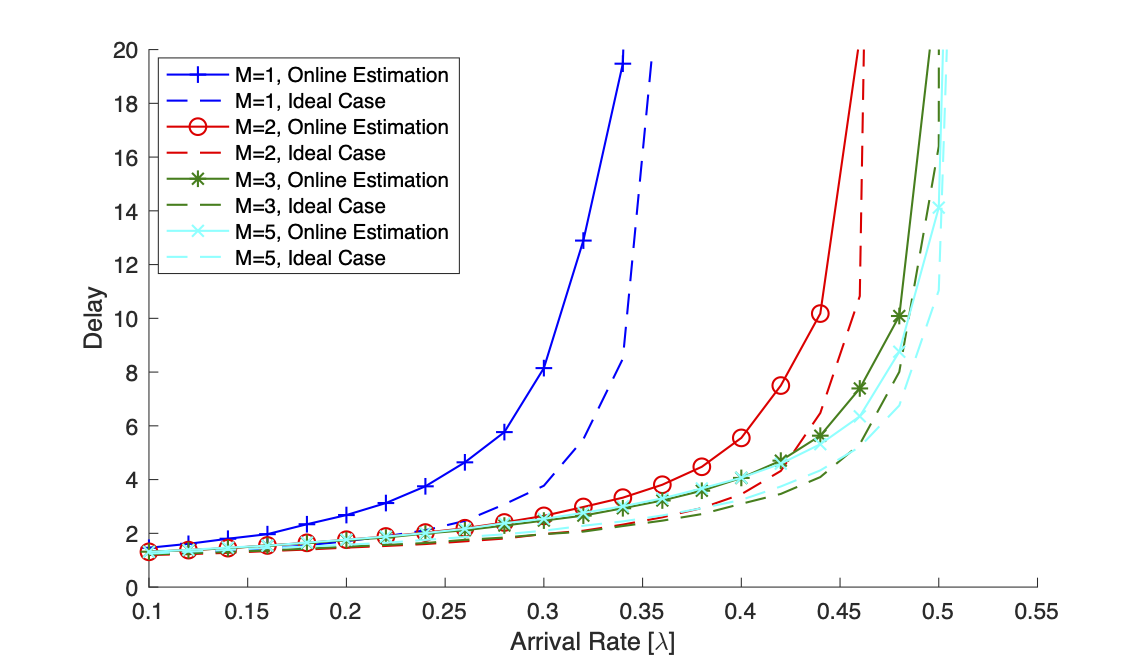}
\caption{Average delay with respect to $\lambda$  when $p_e=0.5$.}\label{fig:fig5}
\end{figure}

\subsection{Impact of SIC Failure}
To evaluation the performance degradation due to SIC failure, Figs. \ref{fig:fig4} and \ref{fig:fig5}  plot the average throughput and delay performance of the proposed algorithm when the probability of SIC failure is given by $p_e=0.5$.
It is worthwhile to mention  that the maximum arrival rate supported by the proposed algorithm coincides with the analytical results of $\hat{S}^*_M$ in Table \ref{table:parameters_sic_fail}.
In spite of the performance degradation due to SIC failure, the results show that SIC is still beneficial for improving throughput compared to the known best achievable throughput of $0.487$ without applying SIC \cite{bert} even for the very restrictive environment of $M=3$ and $p_e=0.5$.
The overall delay performance shows similar tendency as the case of no SIC failure as in Fig. \ref{fig:fig3}.
Because of retransmissions of SIC-failed packets, increasing the access delay is inevitable. Nonetheless, the proposed algorithm optimally adjusts the transmission probabilities during the SIC-enabled collision resolve procedure to minimize the retransmissions due to SIC failure in probabilistic sense.

\section{Conclusion} \label{sec:conclusion}
In this paper, we proposed the online estimation and adaptive transmission algorithm for slotted random access systems.
The proposed algorithm has been designed to fully utilize the SIC capability of the AP by carefully adjusting its transmission via online estimated information.
Thanks to such online estimation and adaptation features, it is universally applicable for general non-Poisson arrivals and also robust to limited SIC capability and SIC failure.


\end{document}